\documentclass[conference]{IEEEtran}
\IEEEoverridecommandlockouts
\usepackage{cite}
\usepackage{amsmath,amssymb,amsfonts}
\usepackage{algorithmic}
\usepackage{graphicx}
\usepackage{caption,subcaption}
\usepackage{multirow}
\usepackage{booktabs}
\usepackage{textcomp}
\usepackage{xcolor}
\usepackage{diagbox}
\usepackage{multirow}
\usepackage{multirow}
\usepackage{graphicx}
\usepackage{colortbl}
\usepackage{pifont}
\usepackage{hhline}
\usepackage{threeparttable}
\usepackage{float}  

\usepackage {subcaption}
\usepackage{graphicx}  
\usepackage{makecell}
\usepackage{comment}

\begin{document}

\title{A 3D Modeling Method for Scattering on Rough Surfaces at the Terahertz Band\label{key}}
\author{\IEEEauthorblockN{Ben Chen\IEEEauthorrefmark{1}, Ke Guan\IEEEauthorrefmark{1}, Danping He\IEEEauthorrefmark{1}, Pengxiang Xie\IEEEauthorrefmark{1}, Zhangdui Zhong\IEEEauthorrefmark{1},
Jianwu Dou\IEEEauthorrefmark{3}, \\Shahid Mumtaz\IEEEauthorrefmark{4}, and Wael Bazzi\IEEEauthorrefmark{5}}

\IEEEauthorblockA{\IEEEauthorrefmark{1}State Key Laboratory of Rail Traffic Control and Safety, Beijing Jiaotong University, 100044, Beijing, China}


\IEEEauthorblockA{\IEEEauthorrefmark{3}State Key Laboratory of Mobile Network and Mobile Multimedia Technology, 518055, Shenzhen, China}

\IEEEauthorblockA{\IEEEauthorrefmark{4}School of Engineering, Nottingham Trent University, UK}

\IEEEauthorblockA{\IEEEauthorrefmark{5}Electrical and Computer Engineering Department, American University in Dubai, Dubai, UAE}
Corresponding Author: Ke Guan (e-mail: kguan@bjtu.edu.cn).



}

\maketitle

\begin{abstract}
The terahertz (THz) band (0.1-10~THz) is widely considered to be a candidate band for the sixth-generation mobile communication technology (6G). However, due to its short wavelength (less than 1~mm), scattering becomes a particularly significant propagation mechanism. In previous studies, we proposed a scattering model to characterize the scattering in THz bands, which can only reconstruct the scattering in the incidence plane. In this paper, a three-dimensional (3D) stochastic model is proposed to characterize the THz scattering on rough surfaces. Then, we reconstruct the scattering on rough surfaces with different shapes and under different incidence angles utilizing the proposed model. Good agreements can be achieved between the proposed model and full-wave simulation results. This stochastic 3D scattering model can be integrated into the standard channel modeling framework to realize more realistic THz channel data for the evaluation of 6G.

\end{abstract}


\section{INTRODUCTION}

With the substantial commercialization of the fifth-generation mobile communication technology (5G), the global communication industry has started the research and exploration of the sixth-generation mobile communication technology (6G) \cite{Terahertz1}. In the future 6G era, data rate, delay, and device connection density will be greatly improved to further support diverse applications of 6G [2, 3], with an expected peak data rate of terabits per second (Tbps) \cite{9794668}.


The THz spectrum (0.1-10~THz) is generally considered the potential band for 6G due to its large bandwidth~\cite{Hybrid} and excellent penetration performance~\cite{Rahaman2020ReviewingTS}. However, when the frequency moves to the THz band, the wavelength in the sub-millimeter range is close to the size of microstructure on rough surfaces~\cite{4380579}. Surfaces that are considered smooth at lower frequency bands can become incredibly rough for THz waves~\cite{9464918}. Therefore, scattering will play a significant role in the propagation at THz bands~\cite{9411143}.

In recent years, numerous research has been carried out on the scattering characteristics of rough surfaces in the THz band. It was pointed out in \cite{Dispersion} that the roughness of the reflected surface was a limiting factor for the error rate performance of non-line-of-sight (NLOS), high data rate THz wireless communication systems. The authors in \cite{Comparison} proposed a method to measure ultra-wideband THz channels on 14 rough surfaces with different materials from 500~GHz to~750 GHz, calculating the path loss including absorption and diffuse scattering on the rough surface of each material. The authors of \cite{A1} studied THz channels from 300~GHz to 310~GHz for slightly rough surfaces, and compared the channel transfer functions among the Rayleigh-Rice (R-R) model, the Beckmann-Kirchhoff (B-K) model, and the improved B-K model. In \cite{Terahertz2}, the roughness of the non-Gaussian surface was calculated and then imported into the 3D ray-tracer to simulate the scattered power. It was observed that in NLOS scenarios, the deviation of the received power becomes more obvious on rougher surfaces. In \cite{Time}, 
the Beckmann-Spizzichino model was found to be insufficient to accurately simulate diffuse scattering in the THz band. The authors also mentioned that diffuse scattering would be an important factor of interference in the THz communication system. Therefore, to better support future THz communication systems, it is significant to study the scattering mechanism in the THz band, and further establish a low-complexity and universal model. However, although the above discussions have measured and analyzed the scattering on rough surfaces in the THz band, an effective model has not been proposed.

In previous work, we obtained the distribution of scattering on rough surfaces with the help of full-wave simulations. Based on the directive scattering (DS) model \cite{degli2007measurement}, we proposed a two-dimensional (2D) model to reconstruct the scattering distribution in the incidence plane \cite{On}. In this paper, we propose a 3D stochastic model to characterize the scattering on rough surfaces in a more comprehensive way. The proposed THz scattering model is basically identical to the simulation results and can effectively reveal the effect of the roughness on the amplitude and spatial distribution of the scattering.

The rest of the paper is organized as follows:
Section II briefly introduces the DS model and summarizes the limitation of the DS model in characterizing the scattering in THz bands. In Section III, the 3D scattering model characterizing the effect of the roughness is proposed. Based on the proposed model, its applicability to rough surfaces with different incidence angles and shapes is explored in Section IV. Finally, the conclusion and future work are given in Section V.

\section{SCATTERING OF ROUGH SURFACES}

\subsection{Full-wave simulations}
To obtain the data of scattering on rough surfaces, in our previous works, rough surfaces are imported into Feko for full-wave simulations \cite{On}.
In this paper, we followed the same configuration as previous studies: a 300~GHz Transverse Magnetic (TM) wave illuminates the rough surface at $45^{\circ}$. The size for the rough surface used for simulations is 50~mm~$\times$~50~mm, its root-mean-square height $\delta$ is 0.5 mm and the correlation length $l$ is 8 mm. Details for the simulation configuration can be found in Table \ref{configuration1}.

\begin{table}[htbp]
	\vspace{-0.2cm}
	\centering
	\caption{Simulation Configuration}
	\renewcommand{\arraystretch}{1.1}
	\begin{tabular}{c|cc}
		\Xhline{0.7mm}
		\multirow{3}[6]{*}[1.7ex]{\makecell[c]{Incident wave}} & Amplitude & 1 V/m \\
		\cline{2-3}          & Frequency & 300 GHz \\
		\cline{2-3}          &  Incidence Angle &  45$^{\circ}$ \\
		\Xhline{0.4mm}
		\multirow{3}[6]{*}[1.7ex]{Rough surface} & Area  & 50 mm $\times$ 50 mm \\
		\cline{2-3}          & Roughness & $\delta$: 0.5 mm, $l$: 8 mm \\
		\cline{2-3}          & Material & Perfect electric conductor (PEC) \\
		\Xhline{0.7mm}
	\end{tabular}%
	\label{configuration1}
	\vspace{-0.4cm}
\end{table}%

\subsection{Directive scattering model}
\begin{figure}[H]
	\setlength{\abovecaptionskip}{-0.2cm}
	\setlength{\belowcaptionskip}{-0.cm}
	\begin{center}
		\vspace{-0.4cm}
		\noindent
		\includegraphics[width=2.3in]{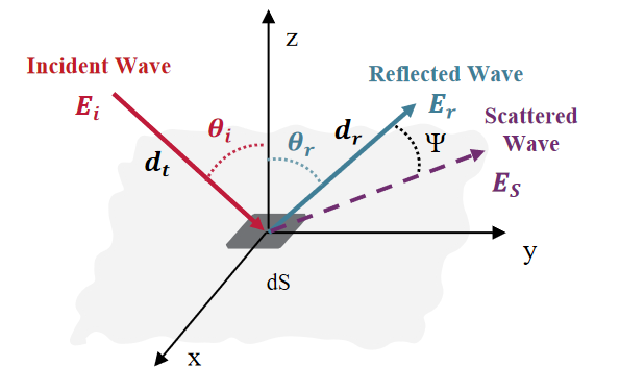}
		\caption{Schematic diagram of the DS model \cite{On}}\label{DS}
	\end{center}
	\vspace{-0.4cm}
\end{figure}

The DS model is widely used to characterize the scattering of targets. It assumes that the scattering lobe is concentrated around the direction of specular reflection. 
The scattered power density of the DS model -- $Pd_{\rm DS}$ -- is given by~\cite{On}:
\begin{equation}
	\vspace{-0.1cm}
	\footnotesize{Pd_{\rm DS}=\frac{|E_{\rm DS}|^2}{\eta}=(\frac{SK}{d_t d_r})^2 \cdot \frac{\rm dS cos\theta_i}{\eta\cdot F_{\alpha_R}} \cdot (\frac{1+\cos(\Psi)}{2})^{\alpha_{R}}}
	\label{DSMODEL}
	\vspace{0.1cm}
\end{equation}
where $E_{\rm DS}$ is the scattered delectric field using the DS model. $S$ is the scattering coefficient, described as $S=|E_s|/|E_i|$. $\eta$ is the free space impedance, $\eta$ = 120$\pi$. As shown in Fig.~\ref{DS}, $E_i$ and $E_s$ are the incident wave and scattered wave, respectively. $d_t$ and $d_r$ represent the distances from the rough surface element $\rm dS$ to the transmitter (Tx) and receiver (Rx), respectively. $K$ is a constant defined as $K = \sqrt{60P_{t}G_{t}}$, where $P_t$ is the transmitted power and $G_t$ is the Tx antenna gain. $\Psi$ is the angle between the scattering direction and the specular reflection direction. $\alpha_{R}$ is the equivalent roughness, which is related to the width of the scattering lobe. The factor $F_{\alpha_R}$ serves as a scaling parameter for normalizing the power scattered by the element \cite{Scattering}.

\subsection{Limitations of the DS model}

\begin{figure}[H]
	\vspace{-0.3cm}
	\centering
	\begin{subfigure}[b]{0.23\textwidth}
		\includegraphics[width=\textwidth]{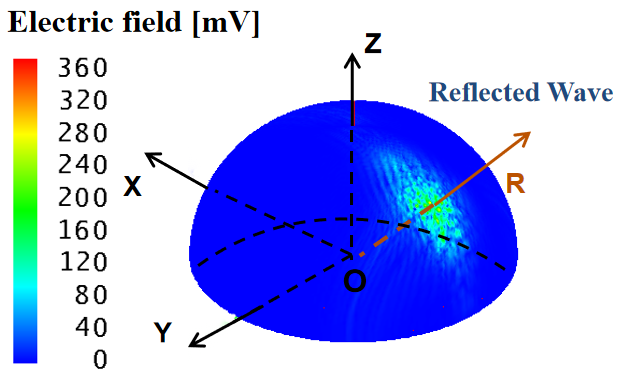}
		\caption{}
		\label{EFEKO}
	\end{subfigure}
	\begin{subfigure}[b]{0.23\textwidth}
		\includegraphics[width=\textwidth]{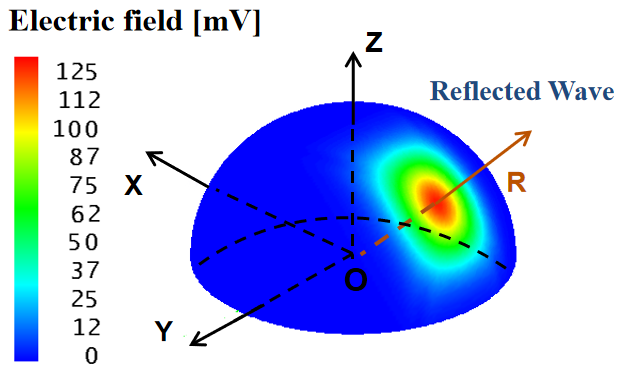}
		\caption{}
		\label{EDS}
	\end{subfigure}
	\caption{Comparison of (a) $E_{\rm Feko}(\theta_s,\phi_s)$ and (b) $E_{\rm DS}(\theta_s,\phi_s)$}
	\label{E}
	\vspace{-0.2cm}
\end{figure}

Based on full-wave simulations, the DS model is used to fit the distribution of scattering on the rough surface. Fig.~\ref{E}~(a) shows the distribution of the scattered electric field on the whole 3D surface by Feko, defined as $E_{\rm Feko}(\theta_s,\phi_s)$, where x, y, and z are the axis of the Cartesian coordinate system, and OR represents the direction of specular reflection. Besides, $\theta_s$ and $\phi_s$ are the zenith angle and the azimuth angle in the spherical coordinate system, respectively. The shape of the scattering lobe is consistent with the assumption of the DS model, which is gathered around the direction of specular reflection. 
Using the DS model in Eq. (\ref{DSMODEL}) and the 3D interpolation method proposed in \cite{A2}, we reconstruct the scattered electric field on the rough surface. The distribution of $E_{\rm DS}(\theta_s,\phi_s)$ is shown in Fig. \ref{E} (b). The pattern of the DS model is an elliptical circle. The scattered electric field decreases from the direction of specular reflection to the surroundings. Comparing Fig.~\ref{E}~(a) with Fig.~\ref{E}~(b), there are obvious limitations of the DS model. The distribution of $E_{\rm Feko}$ is highly random. However, the distribution of $E_{\rm DS}$ is too regular to reflect the randomness caused by the roughness. 


\section{Modeling for the scattering on rough surfaces}

Due to the limitations of the DS model, a stochastic model characterizing the scattering on rough surfaces in THz bands is proposed in this section.

\subsection{Main lobe and non-main lobe}

\begin{figure}[H]
	\vspace{-0.5cm}
	\centering
	\begin{subfigure}[b]{0.24\textwidth}
		\includegraphics[width=\textwidth]{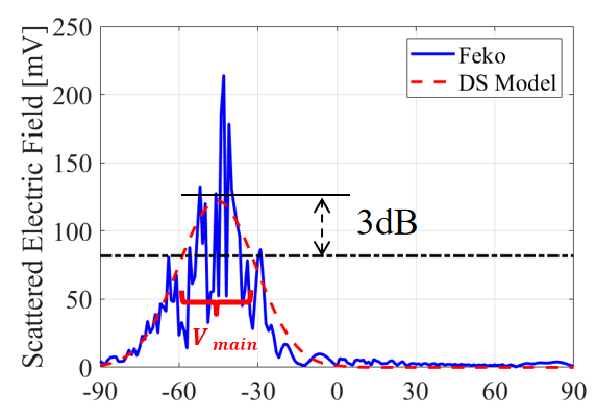}
		\caption{}
		\label{EDS}
	\end{subfigure}
	\begin{subfigure}[b]{0.24\textwidth}
		\includegraphics[width=\textwidth]{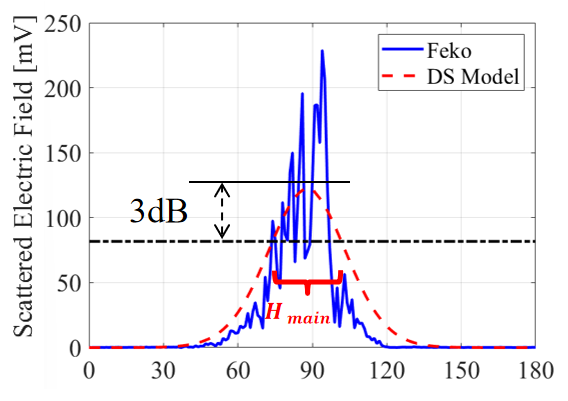}
		\caption{}
		\label{EFEKO}
	\end{subfigure}
	\caption{(a) V-plane and (b) H-plane of the main lobe}
	\label{mainlobe}
	\vspace{-0.3cm}
\end{figure}

As shown in Fig. 2, strong scattering components are mainly concentrated around the specular reflection direction. Therefore, the whole 3D surface is divided into two parts, the main lobe region, and the non-main lobe region. 
To improve the accuracy of modeling, we use the DS model to fit the scattering on two cross-sections (YoR plane and XoR plane) to determine the elliptical main lobe, as shown in Fig.~\ref{mainlobe}.
The YoR plane is defined as the H-plane and the XoR plane is defined as the V-plane. 

 We calculate the 3~dB width of the two DS curves in the V-plane and H-plane, denoted as $V_{main}$ and $H_{main}$, respectively.
If $|\theta_s$-$\theta_r|<V_{main}/2$ and $|\phi_s$-$\phi_r|<H_{main}/2$, the angle~($\theta_s$,$\phi_s$) is located in the main lobe region. $\theta_r$ and $\phi_r$ are the zenith angle and the azimuth angle of specular reflection, respectively. The other region is defined as the non-main lobe region. 

In the main lobe region, $Pd_{\rm rough}$ is added to describe the redistribution due to the roughness. The data for $Pd_{\rm rough}$ is drawn from the difference between the scattered power density by Feko and the DS model, where $Pd_{\rm rough}={|E_{\rm Feko}|^2/\eta}-Pd_{\rm DS}$.
As for the non-main lobe region, the DS model fits well. Thus, $Pd_{\rm DS}$ is used to characterize the scattered power density in this region. Finally, the scattered power density on the whole 3D surface is given as follows:

\vspace{-0.2cm}

\begin{equation}
	\footnotesize{
	\begin{split}
		Pd_{ s}(\theta_{s},\phi_{s}) &= \frac{|E_{s}(\theta_{s},\phi_{s})|^2}{\eta} \\
		&= 		\begin{cases}{}
			Pd_{\rm DS}(\theta_{s},\phi_{s})+Pd_{\rm rough}(\theta_{s},\phi_{s}),&\small{\text{main lobe}}\\
			Pd_{\rm DS}(\theta_{s},\phi_{s}),&\small{\text{non-main lobe}}
		\end{cases}
	\end{split}}
\label{EDSthetaphi}
\end{equation}
\vspace{-0.6cm}

\subsection{The deviation angle}
\begin{figure}[H]
	\setlength{\abovecaptionskip}{0.1cm}
	\setlength{\belowcaptionskip}{-0.cm}
	\begin{center}
		\vspace{-0.3cm}
		\noindent
		\includegraphics[width=2.3in]{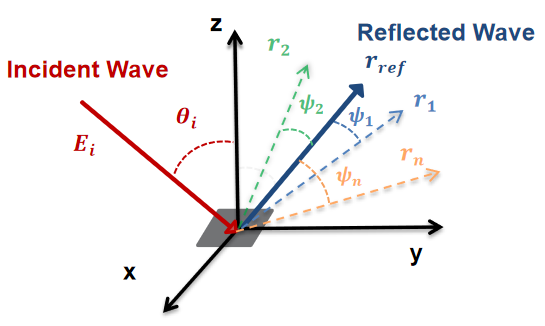}
		\caption{Schematic diagram of $\psi$}\label{deviation}
	\end{center}
	\vspace{-0.4cm}
\end{figure}

To simplify the modeling process, we introduce the concept of deviation angle $\psi$. As shown in Fig. \ref{deviation}, $\psi$ is the angle between $r_{ref}$ and $r_{n}$. The deviation angle $\psi$ is calculated by:
\begin{equation}
\small{\psi_{n}=\arccos [(r_{n}\cdot r_{ref})/(\lvert r_{n}\rvert\cdot|r_{ref}|)]\label{devia}}
\end{equation}
where $r_{n}$ is the unit vector for the $n_{th}$ scattering component, $r_{ref}$ is the unit vector for the direction of specular reflection.

Utilizing Eq.~(\ref{devia}), the scattering angle $\theta_s$ and $\phi_s$ can be transformed into the deviation angle $\psi$. Then, Eq.~(\ref{EDSthetaphi}) can be expressed as:
\vspace{-0.4cm}

\begin{footnotesize}
	\begin{equation}
		Pd_{s}(\psi)=\frac{|E_{s}(\psi)|^2}{\eta}=
		\begin{cases}{}
			Pd_{\rm DS}(\psi)+Pd_{\rm rough}(\psi),&\small{\text{main lobe}}\\
			Pd_{\rm DS}(\psi),&\small{\text{non-main lobe}}
		\end{cases} 
		\label{EDSPSI}
	\end{equation}
\end{footnotesize}

\subsection{Modeling for $Pd_{\rm rough}$}

\begin{figure}[H]
	\setlength{\abovecaptionskip}{0.cm}
	\setlength{\belowcaptionskip}{-0cm}
	\begin{center}
		\vspace{-0.1cm}
		\noindent
		\includegraphics[width=2.1in]{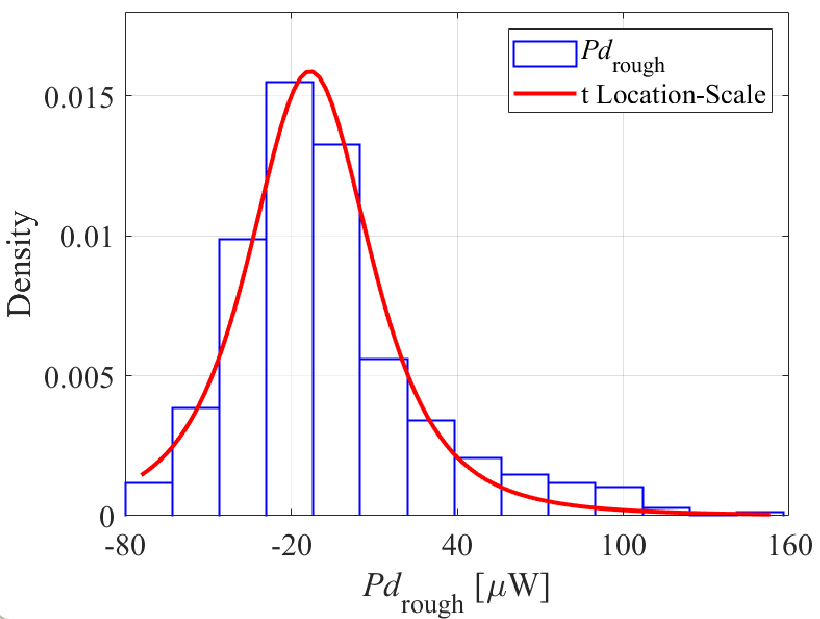}
		\caption{Fitting of $Pd_{\rm rough}$ \cite{On}}\label{Erough}
	\end{center}
	\vspace{-0.6cm}
\end{figure}

The scattered power density in the main lobe region is obtained by adding $Pd_{\rm DS}$ and $Pd_{\rm rough}$. For the rough surface in Section II, the value of $Pd_{\rm rough}$ follows the t Location-Scale distribution, $Pd_{\rm rough}$~$\sim$~T($\mu_t$~=~-12.89, $\sigma_t$ = 8.95, $\nu_t$ = 1.96), as shown in Fig.~\ref{Erough}. The probability density function (PDF) for the t Location-Scale distribution is given by:

\vspace{-0.3cm}
\begin{equation}
	\footnotesize{f(x)=\frac{\Gamma(\frac{\nu_t+1}{2})}{\sigma_t \sqrt{\nu_t\pi}\Gamma(\frac{\nu_t}{2})}\cdot[\frac{\nu_t}{\nu_t+(\frac{x-\mu_t}{\sigma_t})^2}]^\frac{\nu_t+1}{2}}
\end{equation}
where $\mu_t$ is the location parameter, $\sigma_t$ is the scale parameter, and $\nu_t$ is the shape parameter. The t Location-Scale distribution is suitable to fit the data distributions prone to large values.

Because of the power density redistribution caused by the
roughness, some scattering components that are too strong or weak appear in the main lobe. Therefore, the threshold $Pd_{\rm threshold}$ is set to distinguish them. For this case, $Pd_{\rm threshold}$ is equal to the maximum value of $Pd_{\rm rough}$ minus 8~dB. 
Then, $Pd_{\rm rough}$ is divided into $Pd_{\rm rough\_high}$ and $Pd_{\rm rough\_low}$. 

When the power density of the scattering component is greater than $Pd_{\rm threshold}$, it is defined as ${Pd}_{\rm rough\_high}$ and the corresponding deviation angle of the component is $\psi_{\rm high}$.
Conversely, when the power density of the scattering component is smaller than $Pd_{\rm threshold}$, it is defined as $Pd_{\rm rough\_low}$ and the deviation angle of the component is $\psi_{\rm low}$.


\vspace{-0.3cm}

\begin{equation}
\footnotesize{Pd_{\rm rough}(\psi)}=
\begin{cases}{}
	Pd_{\rm rough\_high}(\psi_{\rm high}),&|Pd_{\rm rough}| \geq Pd_{\rm threshold}\\
	Pd_{\rm rough\_low}(\psi_{\rm low}),&|Pd_{\rm rough}| \leq Pd_{\rm threshold}
\end{cases} 
\label{fullloss}
\end{equation}

\begin{figure}[H]
	\setlength{\abovecaptionskip}{0.cm}
	\setlength{\belowcaptionskip}{-0.cm}
	\begin{center}
		\vspace{-0.4cm}
		\noindent
		\includegraphics[width=2.1in]{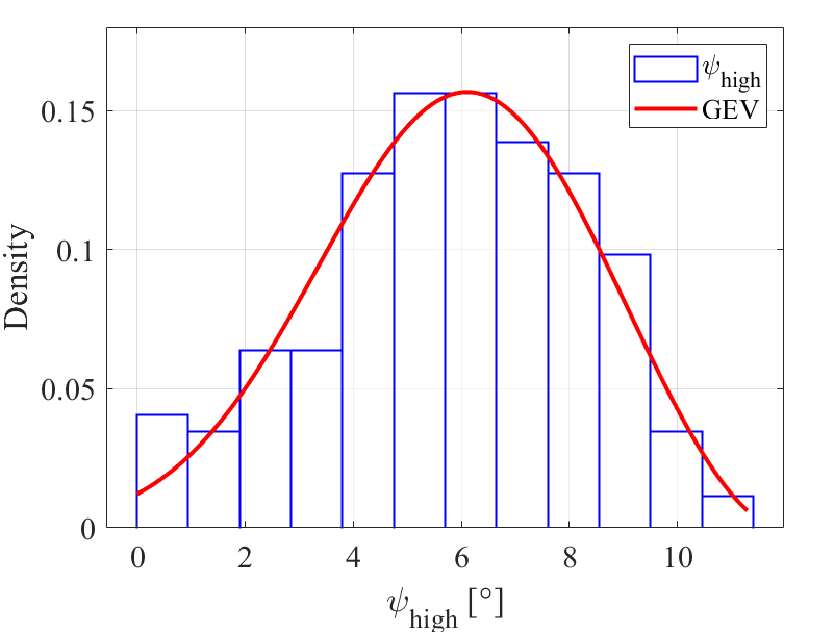}
		\caption{Fitting of $\psi_{\rm high}$}\label{PHI}
	\end{center}
	\vspace{-0.4cm}
\end{figure}

$\psi_{\rm high}$ follows the generalized extreme value (GEV) distribution, $\psi_{\rm high}$ $\sim$ GEV($k_G$ = -0.31, $\sigma_G$ = 2.37, $\mu_G$ = 4.52), as shown in Fig.~\ref{PHI}. 
The PDF for this distribution is given by:

	\vspace{-0.3cm}
\begin{equation}
	\footnotesize{f(x)=\frac{e^{({-(1+k_G\frac{(x-\mu_G)}{\sigma_G})^{-\frac{1}{k_G}}})}}{\sigma_G}\cdot (1+k_G\frac{(x-\mu_G)}{\sigma_G})^{-1-\frac{1}{k_G}}}
\end{equation}
where $\mu_G$ is the location parameter, $\sigma_G$ is the scale parameter, and $k_G$ is the shape parameter ($k_G$ $\neq$ 0).
The GEV distribution typically displays a single peak, the peak of which may appear on the left or right side of the distribution, while the opposite side exhibits the characteristics of a long-tailed distribution. 
$\psi_{\rm low}$ is the empty position from the direction of specular reflection towards the surrounding area. The scattering components are placed at $\psi_{\rm low}$ in descending order of the absolute value of $Pd_{\rm rough\_low}$.

\subsection{Reconstruction of the scattering}

\begin{figure}[H]
	\vspace{-0.2cm}
	\centering
	\begin{subfigure}[b]{0.22\textwidth}
		\includegraphics[width=1.3in]{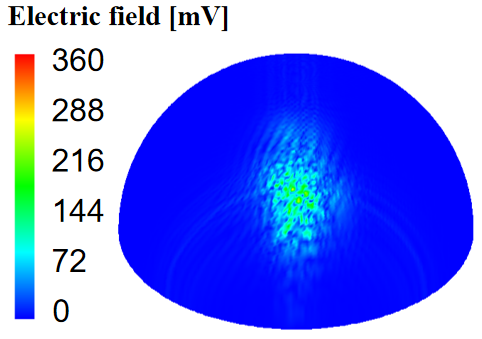}
		\caption{}
		\label{fig:visual_smap_o}
	\end{subfigure}
	\begin{subfigure}[b]{0.22\textwidth}
		\includegraphics[width=1.3in]{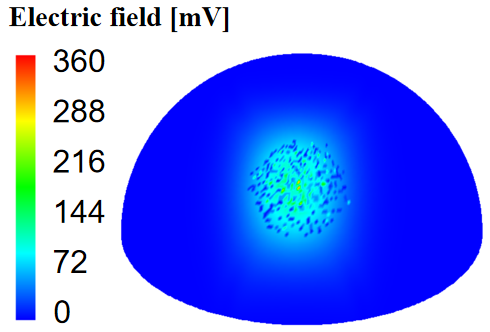}
		\caption{}
		\label{fig:visual_smap_k}
	\end{subfigure}
	\caption{Comparison of the $E_s$ between (a) Feko and the (b) proposed model}
	\label{result}
		\vspace{-0.3cm}
\end{figure}

The proposed model is optimized by superimposing $Pd_{\rm rough}(\psi)$ on the DS model. By considering the amplitude and spatial distribution of $Pd_{\rm rough}$, the scattering distribution characteristics of the rough surface can be characterized more comprehensively and effectively.
Based on the fitting parameters of $Pd_{\rm rough}$ and the DS model, we can reconstruct the distribution of the scattering on the whole 3D surface.
The simulation result by Feko is shown in Fig.~\ref{result}~(a), while Fig.~\ref{result}~(b) shows the scattered electric field $E_s$ reconstructed by the proposed model, which is calculated by Eq. \ref{EDSPSI1}.
In general, the proposed model agrees numerically with the simulation result. 

\vspace{-0.3cm}
\begin{footnotesize}
	\begin{equation}
		{|E_{s}(\psi)|}=
		\begin{cases}{}
			\sqrt{\eta \cdot Pd_{\rm DS}(\psi)+\eta\cdot Pd_{\rm rough}(\psi)},&\small{\text{main lobe}}\\
			{\sqrt{\eta \cdot  Pd_{\rm DS}(\psi)}},&\small{\text{non-main lobe}}
		\end{cases} 
		\label{EDSPSI1}
		\vspace{-0.2cm}
	\end{equation}
\end{footnotesize}

\begin{figure}[H]
	\setlength{\abovecaptionskip}{0.cm}
	\setlength{\belowcaptionskip}{-0.cm}
	\begin{center}
		\vspace{-0.2cm}
		\noindent
		\includegraphics[width=2.2in]{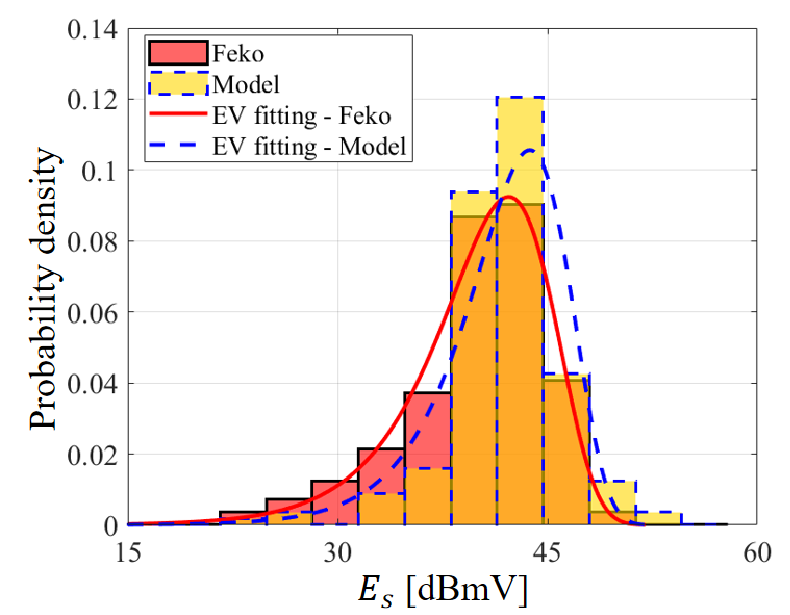}
		\caption{Comparison of $E_s$ between Feko and the proposed model}\label{ERROR1}
	\end{center}
	\vspace{-0.6cm}
\end{figure}

To validate the modeling performance, we compare the scattered electric field $E_s$ in the main lobe region obtained by Feko with that of the proposed model.
Fig.~\ref{ERROR1} illustrates the comparison of the PDF of $E_s$.
As shown in Fig.~\ref{ERROR1}, the red histogram is the value of the scattered electric field by Feko and the yellow histogram is that for the proposed model.
Since the randomness of the generated $Pd_{\rm rough}$ by the t Location-Scale distribution, more values of $E_s$ are at 40-45 dBm. In general, the proposed model is in fundamental consistency with the simulation result.

Moreover, passing the Kolmogorov-Smirnov (KS) test, the extreme value (EV) distribution can well characterize the distribution of $E_s$ in the main lobe region, $E_s$$\sim$EV($\mu_{EV}$, $\sigma_{EV}$), where $\mu_{EV}$ is the position parameter and $\sigma_{EV}$ is the scale parameter.
Table \ref{er1} gives the fitting parameters for $E_s$ by Feko and the proposed model.
$\mu_{EV}$ for Feko and the proposed model are 42.22~dBmV and 43.72~dBmV, respectively. The fitting parameters of Feko and the proposed model are very close, which further indicates that the proposed model can reflect the scattering characteristics of rough surfaces. 

\begin{table}[htbp]
	\centering
	\renewcommand{\arraystretch}{1.1}
	\caption{ Fitting parameters of the EV distribution}
	\renewcommand{\arraystretch}{1.05}
	\begin{tabular}{c|c|c|c}
		\Xhline{0.6mm}
		\multicolumn{2}{c|}{\textbf{\makecell[c]{Feko [dBmV]}}} & \multicolumn{2}{c}{\textbf{Proposed Model [dBmV]}}  \\
		\hline
		\multicolumn{1}{p{1.2cm}<{\centering}|}{\textbf{$\mu_{EV}$}} & \multicolumn{1}{p{1.2cm}<{\centering}|}{\textbf{$\sigma_{EV}$}} & \multicolumn{1}{p{1.2cm}<{\centering}|}{\textbf{$\mu_{EV}$}} & \textbf{$\sigma_{EV}$}  \\
		\hline
		42.22 & 3.97  & 43.72 & 3.48 \\
		\Xhline{0.6mm}
	\end{tabular}%
	\label{er1}%
	\vspace{-0.2cm}
	\vspace{-0.1cm}
\end{table}%
\section{Application of the 3D scattering model}

In the previous section, we introduce the modeling process of the 3D scattering model. To study the universality of the proposed model, in this section, we will utilize it to reconstruct the scattering on rough surfaces with different shapes and under different incidence angles.

\subsection{Different incidence angles}

\subsubsection{Simulation configuration}
In this subsection, incidence angles are set as $15^{\circ}$, $30^{\circ}$, $45^{\circ}$, $60^{\circ}$, and $75^{\circ}$. Other configurations are the same as those in Table I.

\subsubsection{Parameters for modeling}
First of all, the DS model is used to describe the shape of the scattering lobe.
The specific parameters of the DS model are listed in Table \ref{DS1}. As the angle of incidence increases, the $\alpha_R$ on the V-plane fluctuates between 55.16 and 75.89, while the $\alpha_R$ on the H-plane shows a trend of gradual increase from 86.87 to 377.69. This means the scattering component is more concentrated in the incidence plane at large incidence angles.
\begin{table}[htbp]
	\vspace{-0.2cm}
	\centering
	\caption{Fitting parameters of the DS model}
	\renewcommand{\arraystretch}{1.03}
	\begin{tabular}{c|c|c|c|c|c}
		\Xhline{0.6mm}
		\multirow{2}[4]{*}[1.4ex]{\makecell[c]{Incidence \\Angle [$^{\circ}$]}} & \multicolumn{2}{c|}{V-plane } & \multicolumn{3}{c}{H-plane} \\
		\cline{2-6}          & $\alpha_R$ & $S$     & $\alpha_R$ & \multicolumn{2}{c}{$S$} \\
		\Xhline{0.4mm}
		15    & 71.41 & 0.036 & 115.01 & \multicolumn{2}{c}{0.029} \\
		\hline
		30    & 75.89 & 0.032 & 86.87 & \multicolumn{2}{c}{0.026} \\
		\hline
		45    & 57.18 & 0.037 & 103.75 & \multicolumn{2}{c}{0.028} \\
		\hline
		60    & 40.94 & 0.036 & 270.56 & \multicolumn{2}{c}{0.021} \\
		\hline
		75    & 55.16 & 0.047 & 377.69 & \multicolumn{2}{c}{0.012} \\
		\Xhline{0.6mm}
	\end{tabular}%
	\label{DS1}%
	\vspace{-0.2cm}
\end{table}%

Based on the above scattering lobes, $Pd_{\rm rough}(\psi)$ satisfying the t Location-Scale distribution is added to characterize the effect of the roughness. 
The modeling parameters are listed in Table \ref{parameter inci}. 
$V_{main}$ fluctuates between 24$^{\circ}$ and 32$^{\circ}$ as the incidence angle increases. $H_{main}$ decreases from 75$^{\circ}$ to 11$^{\circ}$, when the incidence angle grows from 15$^{\circ}$ to 75$^{\circ}$.

\begin{figure}[htbp]
	\vspace{-0.1cm}
	\centering
	\begin{subfigure}[b]{0.5\textwidth}
		\centering
		\includegraphics[width=\textwidth]{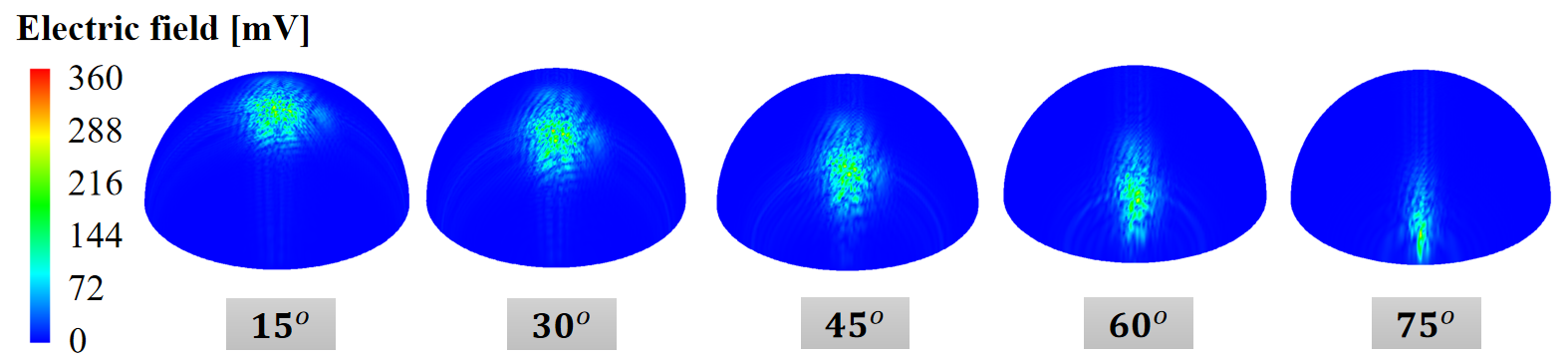}
		
		\vspace{-0.1cm}
		\caption{}
		\label{inci_feko}
	\end{subfigure}
	\hfill
	\begin{subfigure}[b]{0.5\textwidth}
		\centering
		\includegraphics[width=\textwidth]{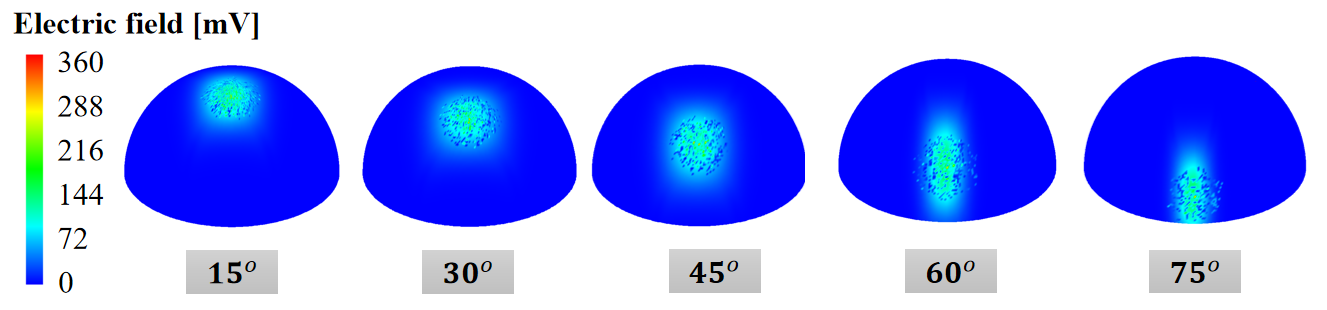}
		
		\vspace{-0.1cm}
		\caption{}
		\label{inci_result}
	\end{subfigure}
	\caption{Comparison of $E_s$ under different incidence angles between (a) Feko and the (b) proposed model}
	\label{inci}
	\vspace{-0.3cm}
\end{figure}

The simulation results by Feko are shown in Fig.~\ref{inci}~(a), while the results of modeling are shown in Fig.~\ref{inci}~(b). 
As the incidence angle becomes larger, the scattering lobe gradually shifts downwards and the shape changes from a circle to a slender line. This trend is well characterized by the proposed model. It agrees very well with the simulation by Feko.


\begin{small}
\begin{table}[htbp]
	\centering
	\caption{The modeling parameters of the proposed model}
	\renewcommand{\arraystretch}{1.03}
	\begin{tabular}{c|c|c|c|c}
		\Xhline{0.7mm}
		\multirow{2}[4]{*}[1.4ex]{\makecell[c]{Incidence \\Angle [$^{\circ}$]}} & \multicolumn{4}{c}{\textbf{3
				dB width}} \\
		\cline{2-5}          & \multicolumn{2}{p{1.8cm}<{\centering}|}{$V_{main}$ [$^{\circ}$]} & \multicolumn{2}{c}{$H_{main}$ [$^{\circ}$]} \\
		\hline
		15    & \multicolumn{2}{c|}{25} & \multicolumn{2}{c}{75} \\
		\hline
		30    & \multicolumn{2}{c|}{24} & \multicolumn{2}{c}{44} \\
		\hline
		45    & \multicolumn{2}{c|}{26} & \multicolumn{2}{c}{28} \\
		\hline
		60    & \multicolumn{2}{c|}{32} & \multicolumn{2}{c}{14} \\
		\hline
		75    & \multicolumn{2}{c|}{28} & \multicolumn{2}{c}{11} \\
		\hline
		\hline
		\multirow{2}[4]{*}[1.4ex]{\makecell[c]{Incidence \\Angle [$^{\circ}$]}} & \multicolumn{4}{c}{\boldmath{$Pd_{\rm rough} \sim T(\mu_t, \sigma_t, \nu_t)$}} \\
		\cline{2-5}          & \multicolumn{1}{p{1.0cm}<{\centering}|}{$\mu_t$}    & \multicolumn{2}{{p{1.0cm}<{\centering}|}}{$\sigma_t$} & $\nu_t$ \\
		\hline
		15    & -10.41 & \multicolumn{2}{c|}{12.07} & 1.48 \\
		\hline
		30    & -7.73 & \multicolumn{2}{c|}{9.07} & 1.29 \\
		\hline
		45    & -12.89 & \multicolumn{2}{c|}{8.95} & 1.96 \\
		\hline
		60    & -16.44 & \multicolumn{2}{c|}{9.99} & 1.89 \\
		\hline
		75    & -20.08 & \multicolumn{2}{c|}{10.18} & 2.25 \\
		
		\hline
		\hline
		\multirow{2}[4]{*}[1.4ex]{\makecell[c]{Incidence \\Angle [$^{\circ}$]}} & \multicolumn{4}{c}{\boldmath{$\psi_{\rm high} \sim {\tiny GEV}(k_G,\sigma_G,\mu_G)$}} \\
		\cline{2-5}          & $k_G$     & \multicolumn{2}{c|}{$\sigma_G$} & $\mu_G$ \\
		\hline
		15    & -0.26 & \multicolumn{2}{c|}{2.61} & 4.63 \\
		\hline
		30    & -0.23 & \multicolumn{2}{c|}{2.64} & 4.57 \\
		\hline
		45    & -0.31 & \multicolumn{2}{c|}{2.37} & 4.52 \\
		\hline
		60    & -0.23 & \multicolumn{2}{c|}{3.17} & 5.65 \\
		\hline
		75    & -0.31 & \multicolumn{2}{c|}{2.02} & 3.37 \\

		\Xhline{0.7mm}
	\end{tabular}%
	\label{parameter inci}%
	\vspace{-0.0cm}
\end{table}%
\end{small}

	\vspace{-0.1cm}

\subsection{Rough surfaces with different shapes}

\subsubsection{Simulation configuration}
In this subsection, the shapes of the rough surface are set as equilateral triangle, square, regular hexagon, and circle, respectively. As mentioned in~\cite{On}, variations in the surface area can significantly affect the far-field scattering. To mitigate this effect, we have ensured that surfaces with different shapes have the same area during simulations, which is still set as $2500$ mm$^2$.


\subsubsection{Parameters for modeling}

First, the DS model is used to fit the H-plane and the V-plane, and the parameters are listed in Table \ref{ds2}. Although the shapes are different, the $\alpha_R$ on the V-plane is always smaller than the $\alpha_R$ on the H-plane, which indicates that the scattering is mainly distributed in the incidence plane.

\begin{table}[htbp]
	\vspace{-0.2cm}
	\centering
	\caption{Fitting parameters of DS model}
	\renewcommand{\arraystretch}{1.02}
	\begin{tabular}{c|c|c|c|c|c}
		\Xhline{0.7mm}
		\multirow{2}[4]{*}[1.4ex]{\makecell[c]{Shape}} & \multicolumn{2}{c|}{V-plane } & \multicolumn{3}{c}{H-plane} \\
		\cline{2-6}          & $\alpha_R$ & $S$     & $\alpha_R$ & \multicolumn{2}{c}{$S$} \\
		\Xhline{0.35mm}
		Triangle    & 35.11 & 0.074 & 314   & \multicolumn{2}{c}{0.015} \\
		\hline
		Square    & 57.18 & 0.037 & 103.76 & \multicolumn{2}{c}{0.028} \\
		\hline
		Hexagon    & 128.02 & 0.025 & 143.34 & \multicolumn{2}{c}{0.033} \\
		\hline
		Circle    & 40.51 & 0.056 & 210.32 & \multicolumn{2}{c}{0.027} \\
		\Xhline{0.7mm}
	\end{tabular}%
	\label{ds2}%
	\vspace{-0.2cm}
\end{table}%


Subsequently, $Pd_{\rm rough}(\psi)$ is added to characterize the influence of microstructures on rough surfaces. Table \ref{3d2} gives the modeling parameters. 
As the shape of the surface changes, $V_{main}$ fluctuates between $18$$^{\circ}$ and $34$$^{\circ}$, while $H_{main}$ fluctuates between $16$$^{\circ}$ and $28$$^{\circ}$.

\begin{figure}[htbp]
	\vspace{-0.4cm}
	\centering
	\begin{subfigure}[b]{0.44\textwidth}
		\centering
		\includegraphics[width=\textwidth]{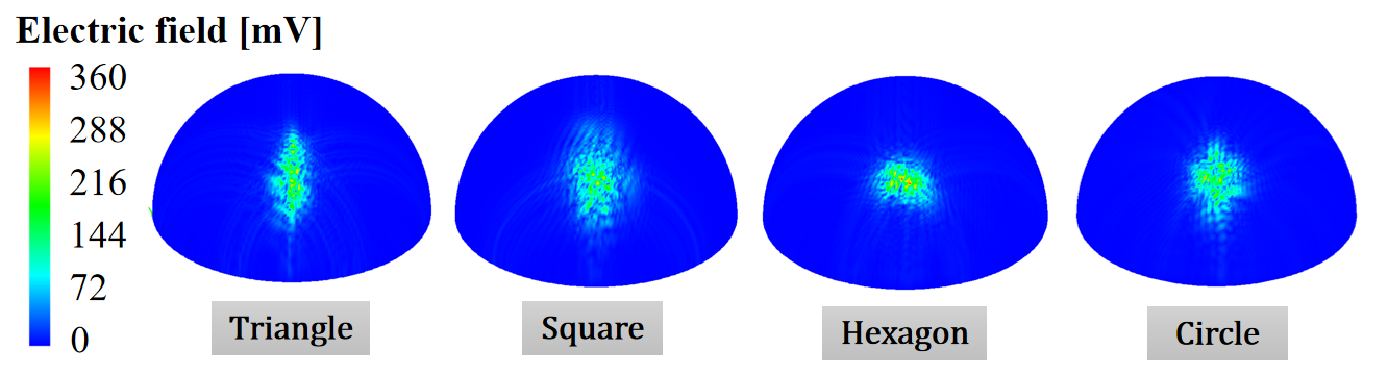}
		
		\vspace{-0.1cm}
		\caption{}
		\label{shape_feko}
	\end{subfigure}
	\hfill
	\begin{subfigure}[b]{0.44\textwidth}
		\centering
		\includegraphics[width=\textwidth]{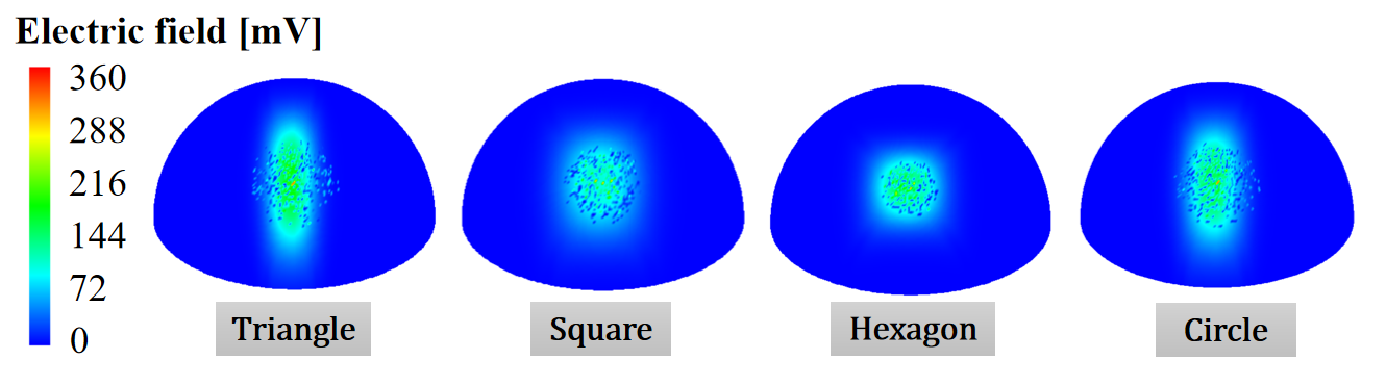}
		
		\vspace{-0.1cm}
		\caption{}
		\label{shpe_result}
	\end{subfigure}
	\caption{Comparison of $E_s$ with different shapes of the rough surface between (a) Feko and the (b) proposed model}
	\label{shape}
	\vspace{-0.6cm}
\end{figure}

The simulation results by Feko are shown in Fig.~\ref{shape}~(a), and the corresponding results of modeling are shown in Fig.~\ref{shape}~(b). 
When the shape of the surface is triangular, square, or circular, the scattering lobe approximates an ellipse, whereas when the surface is hexagonal, it approximates a circle. The variation of the scattering lobe with the shape of the rough surface indicates that the proposed model can characterize the scattering on rough surfaces with different shapes.





\begin{table}[htbp]
		\vspace{-0.1cm}
	\centering
	\caption{The modeling parameters of the proposed model}
	\renewcommand{\arraystretch}{1.03}
	\begin{tabular}{c|c|c|c|c}
		\Xhline{0.7mm}
		\multirow{2}[4]{*}[1.4ex]{\makecell[c]{Shape}} & \multicolumn{4}{c}{\textbf{3
				dB width}} \\
		\cline{2-5}          & \multicolumn{2}{p{1.8cm}<{\centering}|}{$V_{main}$ [$^{\circ}$]} & \multicolumn{2}{c}{$H_{main}$ [$^{\circ}$]} \\
		\hline
		Triangle    & \multicolumn{2}{c|}{34} & \multicolumn{2}{c}{16} \\
		\hline
		Square    & \multicolumn{2}{c|}{26} & \multicolumn{2}{c}{28} \\
		\hline
		Hexagon    & \multicolumn{2}{c|}{18} & \multicolumn{2}{c}{24} \\
		\hline
		Circle    & \multicolumn{2}{c|}{32} & \multicolumn{2}{c}{20} \\
		\hline
		\hline
		\multirow{2}[4]{*}[1.4ex]{\makecell[c]{Shape}} & \multicolumn{4}{c}{\boldmath{$Pd_{\rm rough} \sim T(\mu_t, \sigma_t, \nu_t)$}} \\
		\cline{2-5}          & \multicolumn{1}{p{1.2cm}<{\centering}|}{$\mu_t$}    & \multicolumn{2}{{p{1.2cm}<{\centering}|}}{$\sigma_t$} &\multicolumn{1}{p{1.2cm}<{\centering}}{ $\nu_t$} \\
		\hline
		Triangle    & -23.92 & \multicolumn{2}{c|}{17.73} & 2.52 \\
		\hline
		Square    & -12.89 & \multicolumn{2}{c|}{8.95} & 1.96 \\
		\hline
		Hexagon    & -16.81 & \multicolumn{2}{c|}{15.38} & 2.16 \\
		\hline
		Circle    & -20.81 & \multicolumn{2}{c|}{11.56} & 1.85 \\
		\hline
		\hline
		\multirow{2}[4]{*}[1.4ex]{\makecell[c]{Shape}} & \multicolumn{4}{c}{\boldmath{$\psi_{\rm high} \sim GEV(k_G,\sigma_G,\mu_G)$}} \\
		\cline{2-5}          & $k_G$     & \multicolumn{2}{c|}{$\sigma_G$} & $\mu_G$ \\
		\hline
		Triangle    & -0.27 & \multicolumn{2}{c|}{4.89} & 8.11 \\
		\hline
		Square    & -0.31 & \multicolumn{2}{c|}{2.37} & 4.52 \\
		\hline
		Hexagon    & -0.34 & \multicolumn{2}{c|}{2.61} & 5.48 \\
		\hline
		Circle    & -0.26 & \multicolumn{2}{c|}{2.82} & 5.51 \\
		\Xhline{0.7mm}
	\end{tabular}%
	\label{3d2}%
\end{table}%

\subsection{Evaluation of the proposed model}

To evaluate the proposed model, the scattered electric field $E_s$ in the main lobe region is fitted by the EV distribution for different cases. The fitting parameters of the simulation by Feko and the proposed model are listed in Table~\ref{er}.
For different incidence angles, both the $\mu_{EV}$ of the proposed model and the Feko range between 34~dBmV and 44~dBmV. The $\sigma_{EV}$ are ranged between 3~dBmV and 8~dBmV.
To further evaluate the performance of the proposed model, we calculate the errors of $\mu_{EV}$ and $\sigma_{EV}$ between the proposed model and Feko. 
For different incidence angles, the error of $\mu_{EV}$ ranges from 0.02~dB to 4.41~dB. The error of $\sigma_{EV}$ ranges from 0.49~dB to 4.11~dB. Specifically, when the incidence angle is 30$^{\circ}$, the error of $\mu_{EV}$ reaches a minimum of 0.02~dB.
The error of the fitting parameters increases, as the incidence angle becomes larger. This may be due to the shape of the scattering lobe changing from a circle to a slender line. Therefore, using a width of 3 dB to divide the main lobe region of the proposed model may no longer be appropriate at large incidence angles.

For different shapes of the surface, both the $\mu_{EV}$ of the proposed model and the simulation result range between 37~dBmV and 44~dBmV, the $\sigma_{EV}$ are ranged between 3~dBmV and 7~dBmV. The error of $\mu_{EV}$ ranges from 0.47~dB to 2.34~dB, and the error of $\sigma_{EV}$ ranges from 0.49~dB to 3.19~dB. When the shape of the surface is hexagonal, the error of $\mu_{EV}$ reaches a minimum of 0.47~dB. 

The errors between the proposed model and Feko are acceptable. Therefore, the proposed model can be used to reconstruct the scattering on rough surfaces with different incidence angles and different shapes.
 

\begin{table}[htbp]
		\vspace{-0.2cm}
	\centering
	\renewcommand{\arraystretch}{1.03}
	\caption{Fitting parameters of the EV distribution}
	\begin{tabular}{c|c|c|c|c|c|c}
		\Xhline{0.65mm}
		& \multicolumn{2}{c|}{\textbf{\makecell[c]{Feko\\{[dBmV]}}}} & \multicolumn{2}{c|}{\textbf{\makecell[c]{\vspace{1pt}Proposed Model\\{[dBmV]}}}} & \multicolumn{2}{c}{\textbf{\makecell[c]{Error\\{[dB]}}}} \\
		\hline
		
		\textbf{\makecell[c]{Incidence \\Angle [$^{\circ}$]}} & \textbf{$\mu_{EV}$} & \textbf{$\sigma_{EV}$} & \textbf{$\mu_{EV}$} & \textbf{$\sigma_{EV}$} & \textbf{$\mu_{EV}$} & \textbf{$\sigma_{EV}$} \\
		\hline
		15    & 38.21 & 7.29  & 38.55 & 5.85  & 0.34   & 1.44 \\
		\hline
		30    & 38.11 & 6.11  & 38.13 & 4.92  & 0.02  & 1.19 \\
		\hline
		45    & 42.22 & 3.97  & 43.72 & 3.48  & 1.50   & 0.49 \\
		\hline
		60    & 37.03 & 5.96  & 39.46 & 3.39  & 2.43   & 2.57 \\
		\hline
		75    & 34.44 & 7.44  & 38.85 & 3.33  & 4.41  & 4.11 \\
		\Xhline{0.65mm}
		\textbf{Shape}  & \textbf{$\mu_{EV}$} & \textbf{$\sigma_{EV}$} & \textbf{$\mu_{EV}$} & \textbf{$\sigma_{EV}$} & \textbf{$\mu_{EV}$} & \textbf{$\sigma_{EV}$} \\
		\hline
		Triangle & 39.18 & 5.92  & 41.51 & 3.22  & 2.33   & 2.70 \\
		\hline
		Square & 42.22 & 3.97  & 43.72 & 3.48  & 1.50   & 0.49 \\
		\hline
		Hexagon & 40.01 & 5.11  & 40.48 & 3.61  & 0.47     & 1.50 \\
		\hline
		Circle & 37.97 & 6.56  & 40.31 & 3.37  & 2.34   & 3.19 \\
		\Xhline{0.65mm}
	\end{tabular}%
	\label{er}%
		\vspace{-0.4cm}
\end{table}%

\section{Conclusion and Future Work}

In this paper, we propose a 3D stochastic model to characterize the scattering on rough surfaces in THz bands. Based on the DS model, $Pd_{\rm rough}$ is defined and added to characterize the effect of the roughness. The scattering reconstructed by the proposed model is in good agreement with the full-wave simulation result, which fully represents the scattering characteristics of the rough surface. In addition, the proposed model is capable of modeling scattering on rough surfaces with different incidence angles and different shapes. 
When the incidence angle is smaller than $45^{\circ}$, the errors of the model for both $\mu_{EV}$ and $\sigma_{EV}$ are less than 1.5~dB, suggesting the excellent compliance of modeling.

In the future, we will further model the polarization and phase of scattering on the whole 3D surface. Finally, we expect to establish a scattering model covering multi-dimensional properties such as amplitude, space, phase, and polarization.

	\vspace{-0.2cm}
\section*{Acknowledgment}

This work is supported by the Fundamental Research Funds for the Central Universities 2022JBQY004, the ZTE Corporation and State Key Laboratory of Mobile Network and Mobile Multimedia Technology, the Slovenian Research Agency under grants P2-0016 and J2-4461, and the project (21NRM03 MEWS) which has received funding from the European Partnership on Metrology, co-financed from the European Union's Horizon Europe Research and Innovation Programme and by the Participating States.

\bibliographystyle{IEEEtran}
\bibliographystyle{ACM-Reference-Format}
\bibliography{ref_globalcom}

\begin{thebibliography}{10}
\providecommand{\url}[1]{#1}
\csname url@samestyle\endcsname
\providecommand{\newblock}{\relax}
\providecommand{\bibinfo}[2]{#2}
\providecommand{\BIBentrySTDinterwordspacing}{\spaceskip=0pt\relax}
\providecommand{\BIBentryALTinterwordstretchfactor}{4}
\providecommand{\BIBentryALTinterwordspacing}{\spaceskip=\fontdimen2\font plus
\BIBentryALTinterwordstretchfactor\fontdimen3\font minus
  \fontdimen4\font\relax}
\providecommand{\BIBforeignlanguage}[2]{{%
\expandafter\ifx\csname l@#1\endcsname\relax
\typeout{** WARNING: IEEEtran.bst: No hyphenation pattern has been}%
\typeout{** loaded for the language `#1'. Using the pattern for}%
\typeout{** the default language instead.}%
\else
\language=\csname l@#1\endcsname
\fi
#2}}
\providecommand{\BIBdecl}{\relax}
\BIBdecl

\bibitem{Terahertz1}
D.~Serghiou, M.~Khalily, T.~W.~C. Brown, and R.~Tafazolli, ``Terahertz channel
  propagation phenomena, measurement techniques and modeling for {6G} wireless
  communication applications: A survey, open challenges and future research
  directions,'' \emph{IEEE Communications Surveys \& Tutorials}, vol.~24, no.~4,
  pp. 1957--1996, 2022.

\bibitem{6G}
J.~Wang, C.-X. Wang, J.~Huang, and Y.~Chen, ``{6G THz} propagation channel
  characteristics and modeling: Recent developments and future challenges,''
  \emph{IEEE Communications Magazine}, pp. 1--8, 2022.

\bibitem{9083794}
N.~H. Mahmood, H.~Alves, O.~A. López, M.~Shehab, D.~P.~M. Osorio, and
  M.~Latva-Aho, ``Six key features of machine type communication in {6G},'' in
  \emph{2020 2nd 6G Wireless Summit (6G SUMMIT)}, 2020, pp. 1--5.

\bibitem{9794668}
C.~Han, Y.~Wang, Y.~Li, Y.~Chen, N.~A. Abbasi, T.~Kürner, and A.~F.~Molisch,
  ``Terahertz wireless channels: A holistic survey on measurement, modeling,
  and analysis,'' \emph{IEEE Communications Surveys \& Tutorials}, vol.~24,
  no.~3, pp. 1670--1707, 2022.

\bibitem{Hybrid}
C.~Han, L.~Yan, and J.~Yuan, ``Hybrid beamforming for terahertz wireless
  communications: Challenges, architectures, and open problems,'' \emph{IEEE
  Wireless Communications}, vol.~28, no.~4, pp. 198--204, 2021.

\bibitem{Rahaman2020ReviewingTS}
M.~H. Rahaman, A.~Bandyopadhyay, S.~Pal, and K.~P. Ray, ``Reviewing the scope
  of thz communication and a technology roadmap for implementation,''
  \emph{IETE Technical Review}, vol.~38, pp. 465 -- 478, 2020.

\bibitem{4380579}
R.~Piesiewicz, C.~Jansen, D.~Mittleman, T.~Kleine-Ostmann, M.~Koch, and
  T.~Kurner, ``Scattering analysis for the modeling of thz communication
  systems,'' \emph{IEEE Transactions on Antennas and Propagation}, vol.~55,
  no.~11, pp. 3002--3009, 2007.

\bibitem{9464918}
B.~Ji, Y.~Han, S.~Liu, F.~Tao, G.~Zhang, Z.~Fu, and C.~Li, ``Several key
  technologies for {6G}: Challenges and opportunities,'' \emph{IEEE
  Communications Standards Magazine}, vol.~5, no.~2, pp. 44--51, 2021.

\bibitem{9411143}
M.~Inomata, W.~Yamada, N.~Kuno, M.~Sasaki, K.~Kitao, M.~Nakamura, H.~Ishikawa,
  and Y.~Oda, ``Terahertz propagation characteristics for {6G} mobile
  communication systems,'' in \emph{2021 15th European Conference on Antennas
  and Propagation (EuCAP)}, 2021, pp. 1--5.

\bibitem{Dispersion}
R.~Messenger, K.~Strecker, S.~Ekin, and J.~F. O'Hara, ``Dispersion from diffuse
  reflectors and its effect on terahertz wireless communication performance,''
  \emph{IEEE Transactions on Terahertz Science and Technology}, vol.~11, no.~6,
  pp. 695--703, 2021.

\bibitem{Comparison}
D.~Serghiou, M.~Khalily, S.~Johny, M.~Stanley, I.~Fatadin, T.~W.~C. Brown,
  N.~Ridler, and R.~Tafazolli, ``Comparison of diffuse roughness scattering
  from material reflections at 500-750 {GHz},'' in \emph{2021 15th European
  Conference on Antennas and Propagation (EuCAP)}, 2021, pp. 1--5.

\bibitem{A1}
F.~Sheikh and T.~Kaiser, ``A modified beckmann-kirchhoff scattering model for
  slightly rough surfaces at terahertz frequencies,'' in \emph{2019 IEEE
  International Symposium on Antennas and Propagation and USNC-URSI Radio
  Science Meeting}, 2019, pp. 2079--2080.

\bibitem{Terahertz2}
M.~Alissa, F.~Sheikh, N.~Zarifeh, T.~Kreul, and T.~Kaiser, ``Terahertz wave
  scattering by rough surfaces including higher moments: Ray-tracing
  developments,'' in \emph{2020 IEEE International Conference on Computational
  Electromagnetics (ICCEM)}, 2020, pp. 14--16.

\bibitem{Time}
T.~Attwood, E.~Adams, S.~Freer, A.~J. Vernon, S.~M. Hanham, C.~Constantinou,
  L.~Azpilicueta, and M.~Navarro-Cía, ``Time and frequency analysis of rough
  surface scattering in the thz spectrum,'' in \emph{2021 51st European
  Microwave Conference (EuMC)}, 2022, pp. 237--240.

\bibitem{degli2007measurement}
V.~Degli-Esposti, F.~Fuschini, E.~M. Vitucci, and G.~Falciasecca, ``Measurement
  and modelling of scattering from buildings,'' \emph{IEEE Transactions on
  Antennas and Propagation}, vol.~55, no.~1, pp. 143--153, 2007.

\bibitem{On}
K.~Guan, P.~Xie, D.~He, Z.~Zhong, J.~Dou, and F.~Zhu, ``On the modeling of
  scattering mechanisms of rough surfaces at the terahertz band,'' in
  \emph{Proceedings of the 6th ACM Workshop on Millimeter-Wave and Terahertz
  Networks and Sensing Systems}, 2022, pp. 1--6.

\bibitem{Scattering}
S.~Ju, S.~H.~A. Shah, M.~A. Javed, J.~Li, G.~Palteru, J.~Robin, Y.~Xing,
  O.~Kanhere, and T.~S. Rappaport, ``Scattering mechanisms and modeling for
  terahertz wireless communications,'' in \emph{ICC 2019 - 2019 IEEE
  International Conference on Communications (ICC)}, 2019, pp. 1--7.

\bibitem{A2}
T.~Vasiliadis, A.~Dimitriou, and G.~Sergiadis, ``A novel technique for the
  approximation of {3-D} antenna radiation patterns,'' \emph{IEEE Transactions
  on Antennas and Propagation}, vol.~53, no.~7, pp. 2212--2219, 2005.

\end{thebibliography}

\end{document}